\documentclass[preprint,aps,prb,showpacs,amsmath,amssymb,floatfix]{revtex4}
\usepackage{graphics}
\usepackage{epsf,graphicx,xspace,pstricks,epsfig,psfrag}
\usepackage{color}

\begin{document}

\title{Unconventional superconductivity and interaction induced Fermi surface reconstruction in the two-dimensional Edwards model}
\author{Dai-Ning Cho$^{1}$, Jeroen van den Brink$^{1}$, Holger Fehske$^{2}$, 
Klaus W. Becker$^{3}$, and Steffen Sykora$^{1}$*}


\affiliation{
$^{1}$IFW Dresden, P.O. Box 270116, 01171 Dresden, Germany \\
$^{2}$Institut f{\"u}r Physik, Ernst-Moritz-Arndt-Universit{\"a}t Greifswald, 
D-17487 Greifswald, Germany \\
$^{3}$Institut f{\"u}r Theoretische Physik, Technische Universit{\"a}t Dresden, D-01062 Dresden, Germany \\ \\
*Correspondence to [s.sykora@ifw-dresden.de]
}

{\today}


\begin{abstract}
We study the competition between unconventional superconducting pairing and charge density wave (CDW) formation for the two-dimensional Edwards Hamiltonian at half filling, a very general two-dimensional transport model in which fermionic charge carriers couple to a correlated background medium. Using the projective renormalization method we find 
that a strong renormalization of the original fermionic band causes a new hole-like Fermi surface to emerge near the center of the Brillouin zone, before it eventually gives rise to the formation of a charge density wave. On the new, disconnected  parts of the Fermi surface superconductivity is induced with a sign-changing order parameter.
We discuss these findings in the light of recent experiments on iron-based oxypnictide superconductors.
\end{abstract}



\maketitle

\section*{Introduction}

In a number of superconducting (SC) materials the
pairing interaction is not predominantly mediated by phonons, but rather by  
electron-electron interactions, for instance in the guise of spin fluctuations in the recently discovered extended family of iron-based superconductors \cite{Tsuei2000,Mazin2008,Kuroki2008}. Electron-electron interactions, however, are also driving the metallic ground state towards other long-range ordered states, in particular 
spin\cite{Millis2002,Lester2015}- or charge-density waves\cite{Johannes2008}-\cite{Watanabe2015} (CDWs). A minimal model to effectively describe these interactions considers fermionic charge carriers in the presence of a correlated background that is provided by bosonic modes
in the particle's immediate vicinity which take an active part in the
transport of the fermions\cite{Berciu2009}.
Such a picture is very general with wide
applicability, 
for example to the case of charge transport in high-temperature
SC materials\cite{Wu2011}-\cite{Ghiringhelli2012} where superconductivity appears close to
magnetically ordered phases\cite{Wohlfeld2009}.

The fundamental question arises whether there is a SC state where the Cooper pairing is solely based on electron-electron interaction and in particular whether and how such a phase competes with other ordered states mediated by the same generic background correlations. 
An effective lattice
model which mimics quantum transport in a correlated background
is the Edwards fermion-boson model\cite{Edwards2006},
\begin{equation}
\label{H_real}
\mathcal{H}/ t_b = - \sum\limits_{\langle i,j \rangle} 
c^{\dagger}_{j}c_{i} (b^{\dagger}_{i} + b_{j}) 
- \Lambda \sum\limits_{i} (b^{\dagger}_{i} + 
b_{i}) + \Omega \sum\limits_{i} b^{\dagger}_{i} 
b_{i},
\end{equation}
which is here considered for a 2D square lattice. It describes the hopping of spinless fermions  between nearest-neighbor sites $i$ and $j$ affected by a correlated background medium modelled by bosonic degrees of freedom. Local excitations and quantum fluctuations in the background medium are parametrized by dimensionless parameters $\Omega$ and $\Lambda$, which give the energy cost of a bosonic excitation and the ability of the background to relax, respectively. Originally, the Edwards model was introduced to describe the motion of a spinless particle
in an antiferromagnetic correlated spin background - like a hole in the  $t$-$J$ model.  In this context
the Edwards model is relevant to charge transport in high-temperature superconductors at 
doping levels close to an antiferromagnetically ordered state\cite{Alvermann2007} 
but also in other materials with related 
models with spin degrees of freedom\cite{Edwards2010}. The advantage of the Edwards model is that the correlated spin background is parametrized by bosonic degrees of freedom, which might be represented, for example, by Schwinger bosons. Thus, in the Edwards model the charge carriers are modelled by spinless fermions, whereas the background spin correlations are represented  by bosons. Therefore,  
 the 2D Edwards 
Hamiltonian~\eqref{H_real} also allows  the study of superconductivity using spinless fermions and the spin degrees of freedom are modelled  by bosons in a way described above.  

 Shortly after its 
introduction the model~\eqref{H_real}  was solved numerically for a single particle in 1D  by 
a variational diagonalization technique\cite{Alvermann2007}, and in 2D treated approximatively by the momentum-average approach\cite{Berciu2010}.
The many-particle case has been studied intensively for the 1D system within DMRG, 
where a surprisingly rich 
phase diagram has been found, including metallic repulsive and attractive
Tomonaga-Luttinger-liquid phases, insulating CDW 
states at half-filling\cite{Ejima2009,Sykora2010} and 
one-third-filling\cite{Ejima2014}, and
regions with phase separation\cite{Ejima2012}.  

In this work, we exploit the projective renormalization method (PRM)\cite{Becker2002} to the Edwards model at half-filling and find an intricate interplay between stable superconducting and charge-density wave states, that strongly depends on the excitation energy of the correlated background medium.  The original fermionic band is strongly renormalized by the coupling to the bosonic modes, which can give rise to an entirely new hole-like part of the Fermi surface (FS) close to the center of the Brillouin zone. The  superconducting order parameter has a different sign on the two disconnected parts of the FS that subsequently emerge. Such an interaction induced Fermi surface reconstruction has been observed in recent experiments on oxypnictide superconductors where indeed also sign-changing $s_\pm$ superconductivity is present.

Before discussing the results of our many-particle method for the Edwards model in 2D we start with an appropriate reformulation of Hamiltonian~\eqref{H_real} in such a way that a coupled system of free fermions and bosons is obtained.  Fourier transformation, introduction of fluctuation operators for fermions, and elimination of the linear term of bosonic operators as shown in Ref.~\cite{Sykora2010}  allows the following  decomposition $\mathcal{H} = \mathcal{H}_0 + \mathcal{H}_1$ in momentum space, 
\begin{eqnarray}
\label{H_momentum}
\mathcal{H}_0 &=& \sum\limits_{\mathbf{k}}\varepsilon_{\mathbf{k}} 
\delta(c^{\dagger}_{\mathbf{k}}c_{\mathbf{k}}^{})+ \Omega t_b \sum\limits_{\mathbf{q}} 
b^{\dagger}_{\mathbf{q}}b_{\mathbf{q}} - N \frac{\Omega t_f^2}{4 t_b},  \\ 
\label{H_1}
\mathcal{H}_1 &=& \frac{1}{\sqrt{N}} 
\sum\limits_{\mathbf{k}\mathbf{q}}g_{\mathbf{k}} \left[
b^{\dagger}_{\mathbf{q}} \delta(c^{\dagger}_{\mathbf{k}} 
c_{\mathbf{k}+\mathbf{q}}^{}) + b_{\mathbf{q}} \delta(c^{\dagger}_{\mathbf{k} 
+\mathbf{q}} c_{\mathbf{k}}^{}) \right] ,
\end{eqnarray}
where $\varepsilon_{\mathbf{k}}=-2t(\cos k_{x}a + \cos k_{y}a)$, 
$g_{\mathbf{k}}=-2t_{b} (\cos k_{x}a + \cos k_{y}a)$, and $t = (2t_b /\Omega) [ \Lambda - (1 / N) \sum_{\bf k} (g_{\bf k}/t_b) \langle c_{\bf k}^\dagger c_{\bf k}^{}\rangle ]$. Here $a$ is the 
lattice constant of the 2D square lattice with $N$ sites. Fluctuation operators $\delta(c^{\dagger}_{\mathbf{k}} c_{\mathbf{k}+\mathbf{q}}) = c^{\dagger}_{\mathbf{k}} c_{\mathbf{k}+\mathbf{q}} - \langle c^{\dagger}_{\mathbf{k}} c_{\mathbf{k}+\mathbf{q}} \rangle$ were introduced in order to attribute the mean-field contributions to the free term $\mathcal{H}_0$, which simplifies the solution of the many-body problem by the projective renormalization method (PRM).

\section*{Results}

One of the main aims of our work is to discuss the question whether the Edwards model provides an attractive pairing interaction. If so, we have to clarify its structure in momentum space and in which parameter space the SC phase is stable with respect to other ordered states. To reveal a possible SC pairing mechanism  an approximate BCS-like relation between the SC order parameter and pairing correlation function can be derived from the PRM renormalization equations, $\tilde{\Delta}_{\bf k} \approx (1/N)\sum_{\bf q} V_{{\bf k},{\bf q}} \langle c_{-{\bf q}}c_{\bf q} \rangle$, where $V_{{\bf k},{\bf q}}$ is an approximate analytic expression for the momentum-dependent pairing potential, $V_{{\bf k},{\bf q}}=-4g_{{\bf k}}g_{{\bf k}+{\bf q}}\tilde{\omega}_{\bf q}/[\tilde{\omega}_{\bf q}^2-(\tilde{\varepsilon}_{{\bf k}}-\tilde{\varepsilon}_{{\bf k}+{\bf q}})^2]$. The momentum-dependent quantities $\tilde{\varepsilon}_{\bf k}$ and $\tilde{\omega}_{\bf q}$ are determined self-consistently by the PRM approach and they describe the fully renormalized one-particle energies of the fermions and bosons, respectively (compare Eq.~\eqref{Htilde} from the Methods section).  According to the prefactor $g_{\bf k} = -2t_{b} (\cos k_{x}a + \cos k_{y}a)$, the  pairing potential becomes strongly momentum-dependent with 
a sign-change indicating an unconventional Cooper pairing mechanism.

We have evaluated the PRM renormalization equations in the half-filled band case, i.e. for $N_e / N = 0.5$, where $N_e$ is the number of fermionic particles, and have varied the parameter $\Omega$. The second parameter $\Lambda$ was fixed to a very small value $\Lambda = 0.001$ describing a rather stiff, strongly correlated background which supports the formation of ordered states, as for example the SC and CDW states. 
As a result we found that the 2D Edwards model forms three different ground states. A metallic state at small values of $\Omega$, a SC state in a narrow region for intermediate 
values of the bosonic energy, $\Omega \approx 2.8 \cdots 3.5$, and a quantum phase transition to a CDW state for large values of $\Omega$, which is a characteristic ground state in the limit of large background excitations energies. 

\begin{figure}[t]
\centering
\includegraphics[width=1\columnwidth] {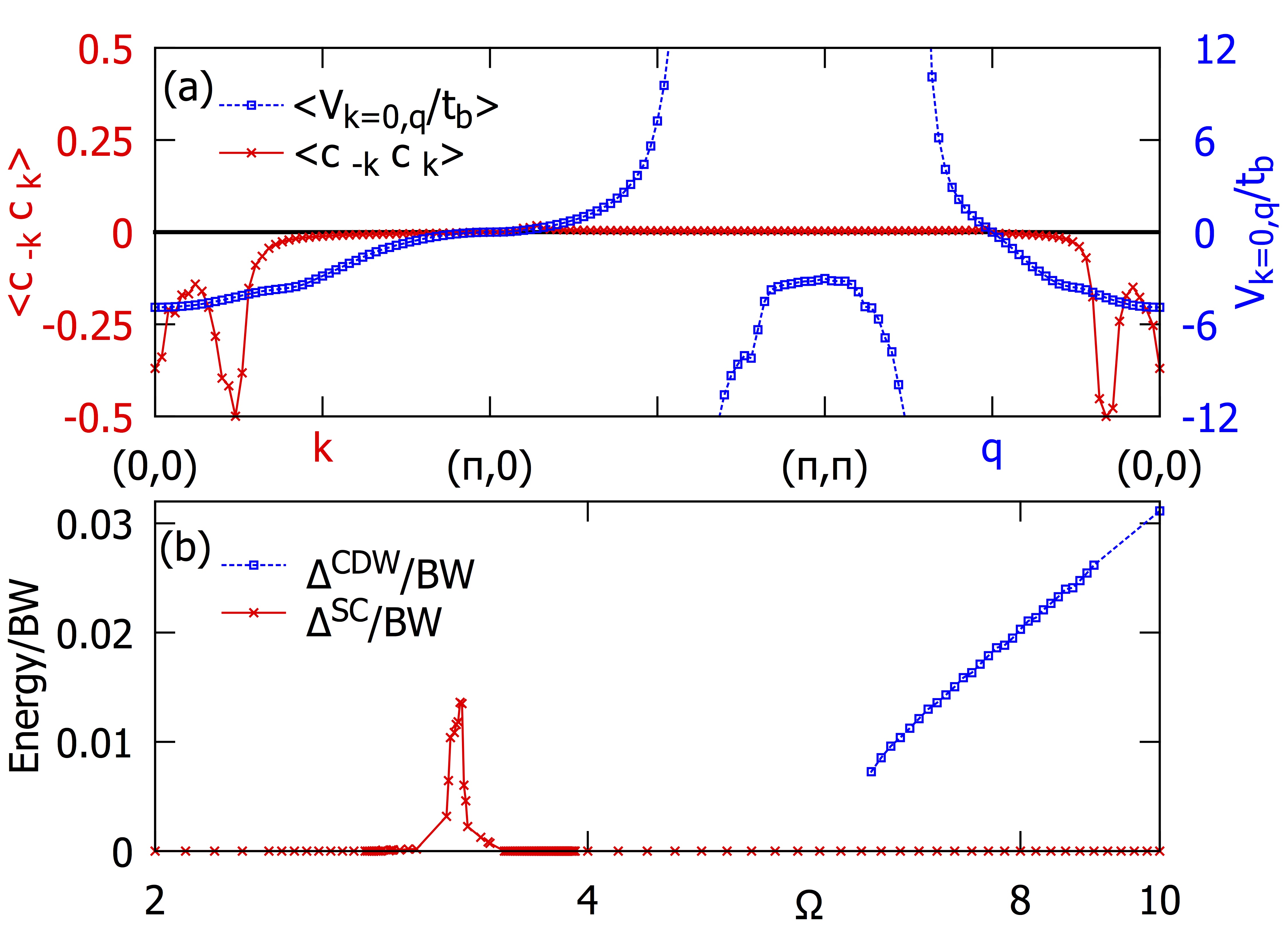}
  \caption{Panel (a): Momentum cuts along the symmetry directions $\Gamma \Rightarrow X \Rightarrow M \Rightarrow \Gamma$ in the SC regime at $\Omega = 3.27$ for the  pairing correlation function $\langle c_{-{\bf k}}c_{\bf k} \rangle$ (left axis) and approximate pairing potential $V_{{\bf k},{\bf q}}/t_{b}$ (right axis)  where ${\bf k}$ is set to ${\bf k} = (0,0)$. The pairing potential is negative in a certain region around the $\Gamma$-point leading to attractive pairing inside the inner hole pocket which is indicated by a non-zero pairing correlation function. Panel (b): Renormalized SC order parameter $\Delta^{SC}$ (red solid line) and charge-density wave order parameter $\Delta^{CDW}$ (blue dashed line) as a function of the bosonic energy $\Omega$. The order parameter values are related to the respective band widths (BW) of the renormalized fermionic bands for the two cases. The lattice grid is 100$\times$100 and the temperature is set to zero.}
\label{fig:OP}
\end{figure}

In  Fig.~\ref{fig:OP}(a) a SC solution at $\Omega = 3.27$ is shown.  Here  the SC pairing correlation function $\langle c_{-{\bf k}}c_{\bf k} \rangle$ and the pairing potential 
$V_{k=0,q}/t_b$ are given 
along momentum cuts in the main symmetry directions. Most notably, we find a pronounced 
tendency towards electron pairing in a certain momentum region around the $\Gamma$ point ${\bf k} = (0,0)$, where also an attractive pairing potential evolves. As is shown further below, the SC state is accompanied by the appearance of a new Fermi surface formed around the $\Gamma$ point in the course of band renormalization, which appears for a specific range of $\Omega$ values. Note that the pairing potential is also negative around $(\pi,\pi)$, where, however, SC pairing is suppressed due to the absence of fermionic low-energy states. The jump in the pairing potential appearing in Fig.~\ref{fig:OP}(a) at momenta where $\tilde{\varepsilon}_{\bf k} \approx \tilde{\omega}_{\bf q}$ is an artifact of the specific perturbative shape of $V_{{\bf k},{\bf q}}$. In the actual calculations this divergency is removed by renormalization contributions up to infinite order in $g_{\bf k}$.

Fig.~\ref{fig:OP}(b) shows solutions of the renormalized SC and CDW order parameters $\Delta^{SC} = \tilde{\Delta}^{SC}_{{\bf k}_F}$ and $\Delta^{CDW} = \tilde{\Delta}^{CDW}_{{\bf k}_F}$ (${\bf k}_F$: Fermi momentum) as a function of $\Omega$. 
Thereby the normalization factors $\rm BW$ in Fig.~\ref{fig:OP}(b) are the band widths of the corresponding renormalized fermionic quasiparticle bands. 

\begin{figure}[t]
\centering
\includegraphics[width=0.7\columnwidth]{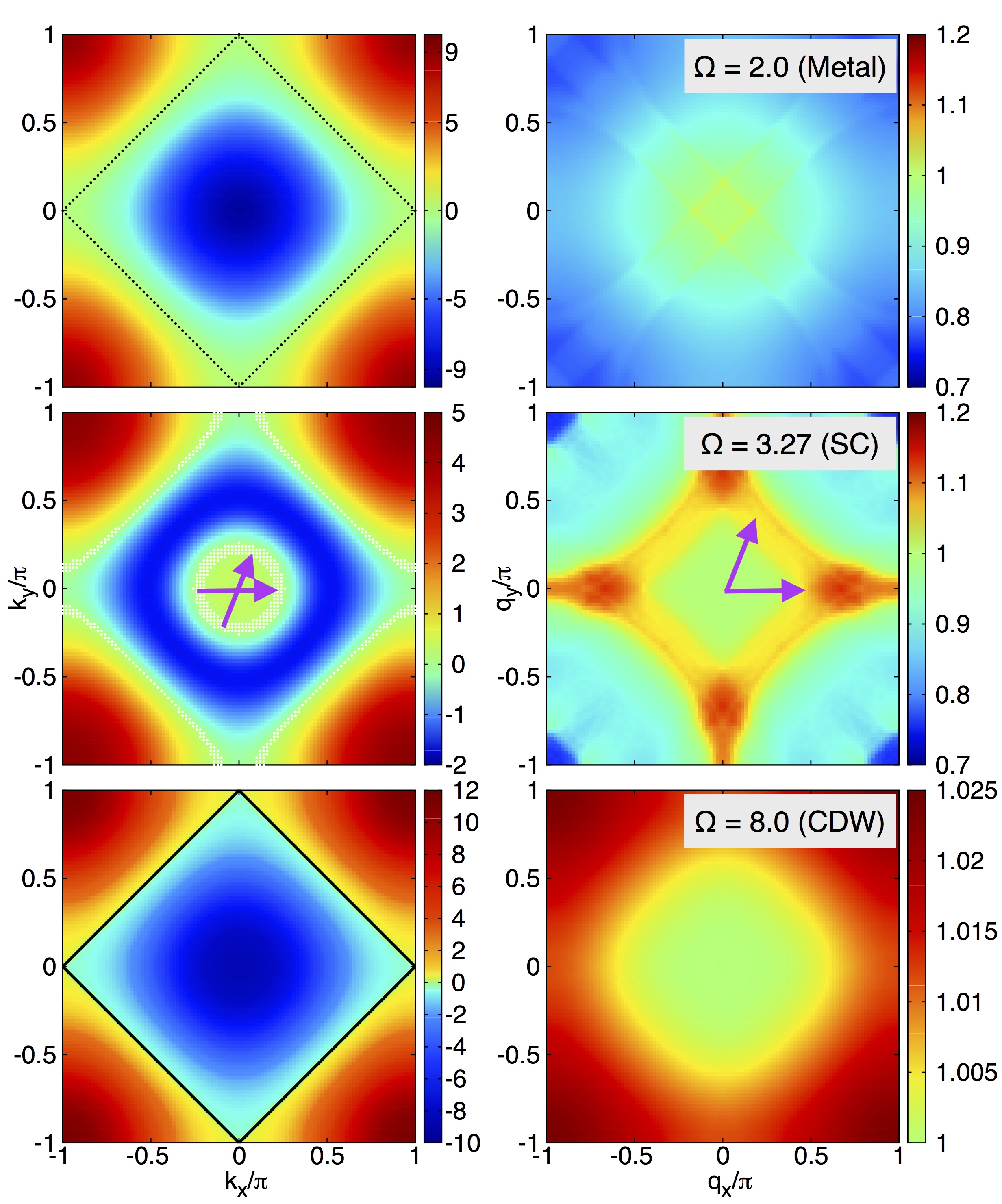}
\caption{Fully renormalized quasiparticle energies
$\tilde{\varepsilon}_{\bf k}/t_{b} =  \varepsilon_{{\bf k},\lambda = 0}/t_{b}$ (left panels) and 
$\tilde{\omega}_{\bf q}/(\Omega t_b) =  \omega_{{\bf q},\lambda = 0}/(\Omega t_b)$ (right panels) of fermions and bosons in the 2D square lattice  Brillouin zone for $\Lambda = 0.001$ and different values of $\Omega$.
$\Omega t_{b}$ is the bare bosonic energy.
 The Fermi surface (black dotted line) and the strong dispersion of $\tilde{\varepsilon}_{\bf k}/t_b$ indicate metallic behavior at $\Omega = 2$. In the SC state ($\Omega = 3.27$)  the momentum dependence of $\tilde{\varepsilon}_{\bf k}/t_{b}$ changes qualitatively, forming a new hole-like Fermi surface around the center of the Brillouin zone. The momentum vectors with 
 $|\tilde{\varepsilon}_{\bf k}| < \Delta^{SC}$ (white dots), where $\Delta^{SC}$ is the SC gap, 
 indicate the position of the respective normal state Fermi surface. It is split into two disconnected parts. Arrows mark representative dominant processes stabilizing the SC state by virtual bosons. 
 For $\Omega = 8$ a CDW state is found. The formation of the CDW gap is  indicated by the black lines, which also encompass the 
 reduced Brillouin zone in the CDW phase.
 The remaining area describes the second quasiparticle band which can be back folded 
 by the CDW ordering vector ${\bf Q} = (\pi,\pi)$ to the reduced Brillouin zone. 
}
\label{fig:epsilon} 
\end{figure}

To  characterize the three phases in more detail, in Fig.~\ref{fig:epsilon} we have considered the fully renormalized one-particle energies $\tilde{\varepsilon}_{\bf k}$ (left panels) and $\tilde{\omega}_{\bf q}$ (right panels) in the entire Brillouin zone. 
%
First, for a small value $\Omega = 2$, we find
typical metallic behavior with a strong dispersion 
$\tilde{\varepsilon}_{\bf k}$, which 
corresponds to the quasiparticle energy. For larger values of $\Omega$, $\tilde{\varepsilon}_{\bf k}$ changes dramatically as can be seen for $\Omega = 3.27$. In this regime, the bosonic energy is in the same order as the renormalized fermionic bandwidth. 
Moreover, $\tilde{\varepsilon}_{\bf k}$ is shifted to values very close to the Fermi level in a certain region in momentum space. 
The strong renormalization leads 
to a new hole-like Fermi surface appearing at momentum $k_F \approx 0.16\pi$, which describes an unoccupied area around the $\Gamma$ point. According to Fig.~\ref{fig:OP}(a) the states inside this region have been detected to be  responsible for the formation of the SC state. However, the low-energy states along the outer Fermi surface are not involved in the SC pairing, because there the pairing potential is zero. For the same reason, combinations of inter-pocket scattering vectors between inner and outer Fermi surface parts do not contribute to the pairing. 
For still larger $\Omega$ inside the CDW regime the SC state is suppressed, due to the absence of a Fermi surface in a momentum region with negative pairing potential. As in the metallic state, the normal state Fermi surface  
runs along the  line where $g_{\bf k}$ is zero, leading to a suppression of the pairing potential.  

For the SC gap we find constant values along the two Fermi surface parts and a characteristic sign change between them. At $\Omega = 3.27$ we have computed the values  $\Delta_1 =1.3 \cdot 10^{-2}$ and  $\Delta_2 = -2.0  \cdot 10^{-4}$ 
 (in terms of the renormalized bandwidth) for the inner hole pocket and the outer part, respectively.  Thus, along the outer Fermi surface the gap value is non-zero but very small, $|\Delta_2| \ll |\Delta_1|$, which suppresses the pairing function $\langle c_{-{\bf k}}c_{{\bf k}} \rangle$ in Fig.~\ref{fig:OP}(a) upon crossing the outer Fermi surface. The sign-change on disconnected parts of the Fermi surface, with $s$-wave pairing on each disjunct part is often denoted as  $s_\pm$  superconductivity in the context of iron-based pnictide superconductors\cite{Johnston2010}. The remarkable feature in the half-filled Edwards model is not as much the $s_\pm$ symmetry of the superconducting order parameter, but rather that the interactions induce a change in the topology of the Fermi surface -- a Lifshitz transition. 
 Such an interaction-induced Fermi-surface reconstruction was recently observed in a prototypical compound 1111-type iron-based oxypnictide superconductor, SmFe$_{0.92}$Co$_{0.08}$AsO, which has a T$_c$ of 55 K. Angle resolved photoemission (ARPES) experiments show that its Fermi surface is reconstructed by the  edges of several bands that are pulled to the Fermi level from the depths of the theoretical band structure\cite{Borisenko2015}. This type of Fermi surface reconstruction is argued to be correlated with the maximally attainable superconducting transition temperature in iron-based superconductors. 
 In particular our observation of a rather high density of states in the vicinity of the inner hole pocket giving rise to the strong Cooper pairing around the $\Gamma$ point is consistent with these ARPES results. 
 %

For larger $\Omega > 4.19$ the hole-like Fermi surface 
disappears again.  For $\Omega > 6.2$ we find a CDW ordered state characterized by a gap in the fermionic quasiparticle spectrum.  A representative CDW solution is shown in Fig.~\ref{fig:epsilon} at $\Omega = 8$ (left panel). The gap opens along the black solid line connecting the momentum vectors $(0,\pm \pi)$ and $(\pm \pi,0)$
and  is reflected in a jump of the color code.
Note that between $\Omega \approx 3.5$ and $\Omega \approx 6.1$ no stable CDW solution was found so that no reliable statement can be given for this 
interval.

Finally, let us characterize the different regimes from the viewpoint of the background correlations. 
In Fig.~\ref{fig:epsilon} (right panel)   
 the renormalized bosonic energy $\tilde{\omega}_{\bf q} /(\Omega t_b)$ 
describes the excitations in the background which have acquired a dispersion  from the coupling 
to the fermions. This renormalization is a consequence of an effective non-local interaction between bosons, which is mediated by higher-order contributions in the fermion-boson coupling.  
As a result the frequency of the boson either increases (hardens) 
or decreases (weakens). 
However, a real soft-mode behavior as it is discussed for instance in the context of structural phase transitions is not observed in the Edwards model in any parameter regime. First, the metallic state $(\Omega= 2.0)$ is characterized 
by a lowering of the effective background energy indicating an 
attractive Fermi liquid. Here the transport is accompanied 
by a 'cloud' of bosonic excitations similar to the large polaron formation
in the presence of phonons. In particular, in the SC state $(\Omega=3.27)$ an interesting structure of the background
energy distribution is found in momentum space. 
The figure shows that $\tilde{\omega}_{\bf q}$ increases inside a Brillouin zone region (green and yellow color), which  is  bounded by scattering vectors ${\bf q}$ that roughly  fulfill the resonance condition 
$|\tilde{\varepsilon}_{\bf k} - \tilde{\varepsilon}_{{\bf k}+{\bf q}}| \approx \tilde{\omega}_{\bf q}$. This leads to a large renormalization with a characteristic change from hardening to softening. Such a behavior is also seen in the jump of the pairing potential in Fig.~\ref{fig:OP}(b) which, however, appears at different momentum vectors due to the 
fixed ${\bf k}$ vector. Inside the boson hardening region we can identify the momentum vectors of the particular bosons which are involved in the SC pairing (represented by two purple arrows in Fig.~\ref{fig:epsilon}). They connect states inside the new inner hole pocket 
where the pairing correlation function is non-zero. 
Finally, in the CDW regime $(\Omega= 8.0)$ the softening area disappears completely and no states are available to form Cooper pairs. Here $\tilde{\omega}_{\bf q} / (\Omega t_b) > 1$ throughout the Brillouin zone and the momentum dependence of the background excitation energy is inverted with respect to the metallic
state emphasizing strongly repulsive particle-particle interaction,
leading immediately to a breakup of any pairing. The hardening 
is a characteristic feature of the CDW phase and has already been found in 1D~\cite{Sykora2010}.

\textbf{Conclusions}. We have studied the interplay between unconventional superconductivity and CDW order within a generic fermion-boson transport model in 2D. In the half-filled case stable SC solutions besides a CDW state were 
found. The SC state is stabilized by an attractive pairing on an additional hole-like Fermi surface around the center of the Brillouin zone which arises through a strong renormalization of the bare fermionic bandstructure.
 The highly unconventional pairing mechanism is due to interaction processes of infinite order in the coupling $g_{\bf k}$ between spinless fermions and background correlations, leading to a momentum-dependent pairing potential which forms a SC state with possible variation of the order parameter phase. Fourier transformation of the $\tilde{\Delta}_{\bf k}$ solution to real space leads to a rather local pairing between next nearest lattice sites. It follows that the character of the SC pairing is influenced by fluctuations of the competing charge ordered state. These results are highly relevant for the currently investigated iron-based oxypnictides.

\section*{Methods}

Following the basic idea of Ref.~\cite{Sykora2010} the interaction part ${\cal H}_1$,
given by Eq.~\eqref{H_1}, is integrated out by a series of unitary 
transformations
starting from large to zero transition energies. Assuming that all transitions 
with energies larger than some
energy cutoff $\lambda$ have already been integrated out, the
transformed Hamiltonian ${\mathcal H}_\lambda$ consists of a part which has the same operator structure as Eqs.~\eqref{H_momentum} and  \eqref{H_1}, but with $\lambda$-dependent parameters, and an additional term of the form $\sum_{\bf k} [ \Delta_{{\bf k},\lambda} c^{\dagger}_{\bf k} 
c^{\dagger}_{-{\bf k}} + \Delta^{*}_{{\bf k},\lambda} 
c_{-{\bf k}}c_{\bf k}]$ allowing for possible 
SC solutions based on an unconventional Cooper 
pairing of spinless fermions.

Evaluating the unitary transformation up to order $g_{\mathbf k}^2$,
discrete renormalization equations for all $\lambda$-dependent parameters 
 are obtained, which  
connect the parameters at cutoff $\lambda$ with those
at  $\lambda - \Delta \lambda$. The discrete renormalization equation for
the SC order parameter function reads
\begin{eqnarray}
\label{Delta}
&&\Delta_{{\bf k},\lambda-\Delta\lambda} - \Delta_{{\bf k},\lambda} 
= -\frac{2 g_{{\bf k}}}{N}  \sum\limits_{{\bf q}}  g_{{\bf k} + {\bf q}} 
\langle c_{-({\bf k}+{\bf q})}c_{{\bf k}+{\bf q}} \rangle \times \nonumber \\
&& \left(\frac{\Theta_{{\bf k},{\bf q},\lambda}^{\Delta \lambda} 
\Theta_{{\bf k} + {\bf q},-{\bf q},\lambda}}{\varepsilon_{{\bf k},\lambda} 
- \varepsilon_{{\bf k}+{\bf q},\lambda} + \omega_{{\bf q},\lambda}} + 
\frac{\Theta_{{\bf k},{\bf q},\lambda} 
\Theta_{{\bf k} + {\bf q},-{\bf q},\lambda}^{\Delta 
\lambda}}{\varepsilon_{{\bf k}+{\bf q},\lambda} - 
\varepsilon_{{\bf k},\lambda} + \omega_{{\bf q},\lambda}} \right),
\end{eqnarray}
where the $\Theta$-functions $\Theta_{{\bf k},{\bf q},\lambda}^{\Delta \lambda} = \Theta_{{\bf k},{\bf q},\lambda} (1 - \Theta_{{\bf k},{\bf q},\lambda - 
\Delta \lambda})$ with $\Theta_{{\bf k},{\bf q},\lambda} = 
\Theta(\lambda - |\varepsilon_{{\bf k},\lambda} -
\varepsilon_{{\bf k} + {\bf q},\lambda} + \omega_{{\bf q},\lambda}|)\,$ restrict the momentum sum to 
excitation energies within a small energy shell $\Delta \lambda$. This allows to apply perturbation theory in each
small renormalization step. However, the overall renormalization 
is far beyond perturbation theory and renormalization processes to infinite
order in the coupling $g_{{\bf k}}$ are taken into account. Furthermore, note 
that in each single renormalization step a factorization approximation 
leads to the appearance of expectation values in the renormalization equations
\cite{Sykora2010}.

The renormalization approach starts by reducing $\lambda$ in steps $\Delta \lambda$ until $\lambda =0$ is reached. This 
is achieved by numerical evaluation of the renormalization equations. Then, all  
transitions from $\mathcal H_{1}$ are used up 
 and the fully renormalized  Hamiltonian $\tilde{\mathcal H} = 
\mathcal{H}_{\lambda =0}$
describes an uncoupled system of bosons and spinless fermions, 
which can be SC ($\tilde{\Delta}_{\bf k} \ne 0$) depending
on the chosen initial parameter set:  The fully renormalized Hamiltonian reads
\begin{eqnarray}
\label{Htilde}
\tilde{\cal H} = \sum_{\bf k} \left[ \tilde{\varepsilon}_{\bf k} 
c^{\dagger}_{\bf k} c^{}_{\bf k} + (\tilde{\Delta}_{\bf k}   c^{\dagger}_{\bf k} 
c^{\dagger}_{-{\bf k}} + \mbox{h.c.}) \right] + \sum_{\bf q} \tilde{\omega}_{\bf q} b^{\dagger}_{\bf q} b_{\bf q}.
\end{eqnarray}

The PRM described in Ref.~\cite{Sykora2010} also allows the calculation of expectation 
values, $\langle \mathcal A \rangle = \langle \tilde{\mathcal A}  \rangle_{\tilde{\mathcal H}}$, where $\tilde{\mathcal A}$ is the fully renormalized quantity using the same set of unitary transformations as for $\tilde{\mathcal H}$. An example is the pairing expectation value  $\langle c_{-{\bf k}}c_{{\bf k}} \rangle$ from Eq.~\eqref{Delta}. In a second step the renormalization procedure starts again 
with the improved expectation
values $\langle c_{-{\bf k}}c_{{\bf k}} \rangle$ by reducing again the 
cutoff from its maximum value  $\bar{\lambda}$ to $\lambda = 0$. After a sufficient number of 
such cycles, the expectation values are
converged and the renormalization equation \eqref{Delta} is solved self-consistently.
Besides SC order we also take into account a possible CDW order (not included in Eq.~\eqref{Htilde}), a phase that is expected to be present at half-filling\cite{Ejima2009,Sykora2010}.

\section*{Author contributions} 
S.S. and H.F. initiated the project. D.N.C. and S.S. developed the theory. D.N.C. performed 
the numerical calculations. S.S., H.F., K.W.B., and J.v.d.B. wrote the main manuscript text and D.N.C. and S.S. prepared the figures. All authors discussed the results and reviewed the
manuscript.

\section*{Acknowledgements} 
We would like to thank S. Ejima, G. Hager, K. Koepernick, and D. Efremov for helpful discussions. This work was supported by Deutsche Forschungsgemeinschaft through the Collaborative Reserach Center SFB 1143, and SFB 652 B5 (H.F.).

\newpage

\begin{thebibliography}{1}
\expandafter\ifx\csname url\endcsname\relax
  \def\url#1{\texttt{#1}}\fi
\expandafter\ifx\csname urlprefix\endcsname\relax\def\urlprefix{URL }\fi
\providecommand{\bibinfo}[2]{#2}
\providecommand{\eprint}[2][]{\url{#2}}

\bibitem{Mazin2008}
\bibinfo{author}{Mazin, I.~I.}, \bibinfo{author}{Singh, D.~J.},  \bibinfo{author}{Johannes, M.~D.} \&
  \bibinfo{author}{Du, M.~H.}
\newblock \bibinfo{title}{Unconventional Superconductivity with a Sign Reversal in the Order Parameter of LaFeAsO$_{1-x}$F$_x$}.
\newblock \emph{\bibinfo{journal}{Phys. Rev. Lett.}}
  \textbf{\bibinfo{volume}{101}}, \bibinfo{pages}{057003}
  (\bibinfo{year}{2008}).

\bibitem{Kuroki2008}
\bibinfo{author}{Kuroki, K.}, \bibinfo{author}{et~al.}
\newblock \bibinfo{title}{Unconventional Pairing Originating from the Disconnected Fermi Surfaces of Superconducting LaFeAsO$_{1-x}$F$_x$}.
\newblock \emph{\bibinfo{journal}{Phys. Rev. Lett.}}
  \textbf{\bibinfo{volume}{101}}, \bibinfo{pages}{087004}
  (\bibinfo{year}{2008}).
  
\bibitem{Tsuei2000}
\bibinfo{author}{Tsuei, C.~C.} \& \bibinfo{author}{Kirtley, J.~R.}
\newblock \bibinfo{title}{Pairing symmetry in cuprate superconductors}.
\newblock \emph{\bibinfo{journal}{Rev. Mod. Phys.}}
  \textbf{\bibinfo{volume}{72}}, \bibinfo{pages}{969}
  (\bibinfo{year}{2000}).

\bibitem{Millis2002}
\bibinfo{author}{Millis, A.~J.}, \bibinfo{author}{Schofield, A.~J.},  \bibinfo{author}{Lonzarich, G.~G.} \&
  \bibinfo{author}{Grigeria, S.~A.}
\newblock \bibinfo{title}{Metamagnetic Quantum Criticality in Metals}.
\newblock \emph{\bibinfo{journal}{Phys. Rev. Lett.}}
  \textbf{\bibinfo{volume}{88}}, \bibinfo{pages}{217204}
  (\bibinfo{year}{2002}).

\bibitem{Lester2015}
\bibinfo{author}{Lester, C.}, \bibinfo{author}{et~al.}
\newblock \bibinfo{title}{Field-tunable spin-density-wave phases in Sr$_3$Ru$_2$O$_7$}.
\newblock \emph{\bibinfo{journal}{Nature Materials}}
  \textbf{\bibinfo{volume}{14}}, \bibinfo{pages}{373}
  (\bibinfo{year}{2015}).

\bibitem{Johannes2008}
\bibinfo{author}{Johannes, M.~D.} \& \bibinfo{author}{Mazin, I.~I.}
\newblock \bibinfo{title}{Fermi surface nesting and the origin of charge density waves in metals}.
\newblock \emph{\bibinfo{journal}{Phys. Rev. B}}
  \textbf{\bibinfo{volume}{77}}, \bibinfo{pages}{165135}
  (\bibinfo{year}{2008}).

\bibitem{Hellmann2012}
\bibinfo{author}{Hellmann, S.}, \bibinfo{author}{et~al.}
\newblock \bibinfo{title}{Time-domain classification of charge-density-wave insulators}.
\newblock \emph{\bibinfo{journal}{Nature Communications}}
  \textbf{\bibinfo{volume}{3}}, \bibinfo{pages}{1069}
  (\bibinfo{year}{2012}).

\bibitem{Watanabe2015}
\bibinfo{author}{Watanabe, H.}, \bibinfo{author}{Seki, K.} \&
  \bibinfo{author}{Yunoki, S.}
\newblock \bibinfo{title}{Charge-density wave induced by combined electron-electron and electron-phonon interactions in 1$T$-TiSe$_2$: A variational Monte Carlo study}.
\newblock \emph{\bibinfo{journal}{Phys. Rev. B}}
  \textbf{\bibinfo{volume}{91}}, \bibinfo{pages}{205135}
  (\bibinfo{year}{2015}).

\bibitem{Berciu2009}
\bibinfo{author}{Berciu, M.}
\newblock \bibinfo{title}{Viewpoint: Challenging a hole to move through an ordered insulator}.
\newblock \emph{\bibinfo{journal}{Physics}}
  \textbf{\bibinfo{volume}{2}}, \bibinfo{pages}{55}
  (\bibinfo{year}{2009}).

\bibitem{Wu2011}
\bibinfo{author}{Wu, T.}, \bibinfo{author}{et~al.}
\newblock \bibinfo{title}{Magnetic-field-induced charge-stripe order in the high-temperature superconductor YBa$_2$Cu$_3$O$_y$}.
\newblock \emph{\bibinfo{journal}{Nature}}
  \textbf{\bibinfo{volume}{477}}, \bibinfo{pages}{191}
  (\bibinfo{year}{2011}).

\bibitem{Chang2012}
\bibinfo{author}{Chang, J.}, \bibinfo{author}{et~al.}
\newblock \bibinfo{title}{Direct observation of competition between superconductivity and charge density wave order in YBa$_2$Cu$_3$O$_{6.67}$}.
\newblock \emph{\bibinfo{journal}{Nature Physics}}
  \textbf{\bibinfo{volume}{8}}, \bibinfo{pages}{871}
  (\bibinfo{year}{2012}).

\bibitem{Ghiringhelli2012}
\bibinfo{author}{Ghiringhelli, G.}, \bibinfo{author}{et~al.}
\newblock \bibinfo{title}{Long-Range Incommensurate Charge Fluctuations in (Y,Nd)Ba$_2$Cu$_3$O$_{6+x}$}.
\newblock \emph{\bibinfo{journal}{Science}}
  \textbf{\bibinfo{volume}{337}}, \bibinfo{pages}{821}
  (\bibinfo{year}{2012}).

\bibitem{Wohlfeld2009}
\bibinfo{author}{Wohlfeld, K.}, \bibinfo{author}{Oles, A.~M.} \&
  \bibinfo{author}{Horsch, P.}
\newblock \bibinfo{title}{Orbitally induced string formation in the spin-orbital polarons}.
\newblock \emph{\bibinfo{journal}{Phys. Rev. B}}
  \textbf{\bibinfo{volume}{79}}, \bibinfo{pages}{224433}
  (\bibinfo{year}{2009}).

\bibitem{Edwards2006}
\bibinfo{author}{Edwards, D.~M.}
\newblock \bibinfo{title}{A quantum phase transition in a model with boson-controlled hopping}.
\newblock \emph{\bibinfo{journal}{Physica B}}
  \textbf{\bibinfo{volume}{378-380}}, \bibinfo{pages}{133}
  (\bibinfo{year}{2006}).

\bibitem{Alvermann2007}
\bibinfo{author}{Alvermann, A.}, \bibinfo{author}{Edwards, D.~M.} \&
  \bibinfo{author}{Fehske, H.}
\newblock \bibinfo{title}{Boson-Controlled Quantum Transport}.
\newblock \emph{\bibinfo{journal}{Phys. Rev. Lett.}}
  \textbf{\bibinfo{volume}{98}}, \bibinfo{pages}{056602}
  (\bibinfo{year}{2007}).
  
\bibitem{Edwards2010}
\bibinfo{author}{Edwards, D.~M.}, \bibinfo{author}{Ejima, S.}, \bibinfo{author}{Alvermann, A.} \&
  \bibinfo{author}{Fehske, H.}
\newblock \bibinfo{title}{A Green's function decoupling scheme for the Edwards fermion-boson model}.
\newblock \emph{\bibinfo{journal}{Journal of Physics: Condensed Matter}}
  \textbf{\bibinfo{volume}{22}}, \bibinfo{pages}{435601}
  (\bibinfo{year}{2010}).

\bibitem{Berciu2010}
\bibinfo{author}{Berciu, M.} \& \bibinfo{author}{Fehske, H..}
\newblock \bibinfo{title}{Momentum average approximation for models with boson-modulated hopping: Role of closed loops in the dynamical generation of a finite quasiparticle mass}.
\newblock \emph{\bibinfo{journal}{Phys. Rev. B}}
  \textbf{\bibinfo{volume}{82}}, \bibinfo{pages}{085116}
  (\bibinfo{year}{2010}).

\bibitem{Ejima2009}
\bibinfo{author}{Ejima, S.}, \bibinfo{author}{Hager, G.} \&
  \bibinfo{author}{Fehske, H.}
\newblock \bibinfo{title}{Quantum Phase Transition in a 1D Transport Model with Boson-Affected Hopping: Luttinger Liquid versus Charge-Density-Wave Behavior}.
\newblock \emph{\bibinfo{journal}{Phys. Rev. Lett.}}
  \textbf{\bibinfo{volume}{102}}, \bibinfo{pages}{106404}
  (\bibinfo{year}{2009}).

\bibitem{Sykora2010}
\bibinfo{author}{Sykora, S.}, \bibinfo{author}{Becker, K.~W.} \&
  \bibinfo{author}{Fehske, H.}
\newblock \bibinfo{title}{Charge-density-wave formation in a half-filled fermion-boson transport model: A projective renormalization approach}.
\newblock \emph{\bibinfo{journal}{Phys. Rev. B}}
  \textbf{\bibinfo{volume}{81}}, \bibinfo{pages}{195127}
  (\bibinfo{year}{2010}).

\bibitem{Ejima2014}
\bibinfo{author}{Ejima, S.} \& \bibinfo{author}{Fehske, H.}
\newblock \bibinfo{title}{Charge-Density-Wave Formation in the Edwards Fermion-Boson Model at One-Third Band Filling}.
\newblock \emph{\bibinfo{journal}{JPS Conf. Proc.}}
  \textbf{\bibinfo{volume}{3}}, \bibinfo{pages}{013006}
  (\bibinfo{year}{2014}).

\bibitem{Ejima2012}
\bibinfo{author}{Ejima, S.}, \bibinfo{author}{Sykora, S.},  \bibinfo{author}{Becker, K.~W.} \&
  \bibinfo{author}{Fehske, H.}
\newblock \bibinfo{title}{Phase separation in the Edwards model}.
\newblock \emph{\bibinfo{journal}{Phys. Rev. B}}
  \textbf{\bibinfo{volume}{86}}, \bibinfo{pages}{155149}
  (\bibinfo{year}{2012}).
  
\bibitem{Becker2002}
\bibinfo{author}{Becker, K.~W.}, \bibinfo{author}{H\"ubsch, A.} \& 
\bibinfo{author}{Sommer, T.}
\newblock \bibinfo{title}{Renormalization approach to many-particle systems.}
\newblock \emph{\bibinfo{journal}{Phys. Rev. B}}
  \textbf{\bibinfo{volume}{66}}, \bibinfo{pages}{235115}
  (\bibinfo{year}{2002}).
  
\bibitem{Johnston2010}
\bibinfo{author}{Johnston, D.~C.}
\newblock \bibinfo{title}{The puzzle of high temperature superconductivity in layered iron pnictides and chalcogenides}.
\newblock \emph{\bibinfo{journal}{Adv. Phys.}}
  \textbf{\bibinfo{volume}{59}}, \bibinfo{pages}{803}
  (\bibinfo{year}{2010}).
  
\bibitem{Borisenko2015}
\bibinfo{author}{Charnukha, A.}, \bibinfo{author}{et~al.}
\newblock \bibinfo{title}{Interaction-induced singular Fermi surface in a high-temperature oxypnictide superconductor}.
\newblock \emph{\bibinfo{journal}{Sci. Rep.}}
  \textbf{\bibinfo{volume}{5}}, \bibinfo{pages}{10392}
  (\bibinfo{year}{2015}).








\end{thebibliography}
\end{document}